\documentclass[12pt]{article}
\usepackage{amssymb,amsmath,amsthm}
\usepackage{latexsym}

\newcommand{\beg}{\begin{equation}}
\newcommand{\ene}{\end{equation}}

\begin{document}
\title{
\textsc{\bf Quantum Information in Space and Time} 
 \footnote{Lecture delivered at
 the 4th
 International Conference on Quantum Information held at Meijo
 University, Japan, Feb. 27 - Mar.1, 2001 and at
 ISI Workshop  on Quantum Computer Theory
 held at the ISI Foundation in Torino, Italy,  June 17 - 30, 2001.}
\\
$~$\\}
\author{
\textsf{Igor V. Volovich}
\\
\emph{Steklov Mathematical Institute}\\
\emph {Russian Academy of Sciences}\\
\emph{Gubkin St. 8, 117866, GSP-1, Moscow, Russia}\\
\emph{e-mail: volovich@mi.ras.ru}
}

\date {~}
\maketitle
\thispagestyle{empty}

\begin{abstract}
Many important results in modern quantum information theory
have been obtained for an idealized situation 
when the spacetime dependence
of quantum phenomena is neglected.
However the transmission and processing of (quantum) 
information is a physical process
in spacetime. Therefore such basic notions in quantum information theory
as the notions of  composite systems, entangled
states and  the channel should be formulated
in space and time.
 We emphasize the importance of 
the investigation of quantum information  in space and time.
Entangled states  
in space and time are considered. A modification of
Bell`s equation which includes the spacetime variables is suggested.
A general relation between quantum theory
and  theory of classical stochastic processes
is proposed. It expresses the condition of local realism
in the form of a {\it noncommutative spectral theorem}.
Applications of this relation to the security 
of quantum key distribution
in quantum cryptography are considered.
\end{abstract}
\newpage

\tableofcontents

%%%%%%%%%%%%%%%%%%%%%%%%%%%%%%%%%%%%%%%%%%%%%%%%%%%%
\section{Introduction}
\label{intro}

Recent remarkable experimental and theoretical results have shown that
quantum effects can provide qualitatively new forms of
communication and computation, sometimes
more powerful then the classical ones. Interesting and important results
obtained in quantum computing, teleportation and cryptography
are based on the  investigation of basic properties of quantum
mechanics. Especially important are  properties of nonfactorized
entangled states
which were named by Schrodinger as the most characteristic 
feature of quantum mechanics.

Modern quantum information theory is built  on 
 ideas of
classical information theory of C. Shannon and on
the notions of von Neumann quantum mechanical entropy 
and of entangled states as formulated by J. Bell, see
~\cite{QI,Ohy,Fuc} for  recent discussions.
The spacetime dependence is not explicitly indicated
in this approach. As a result,  
many important achievements  in modern quantum information theory
have been obtained for an idealized situation 
when the spacetime dependence
of quantum phenomena is neglected. 

We emphasize the importance of 
the investigation of quantum information effects in space and time.
\footnote{The importance of 
the investigation of quantum information effects in space and time
and especially the role
of relativistic invariance in classical and quantum information theory
was stressed in the talk by the author  
at the First International Conference
on Quantum Information which was held at Meijo University, Japan,
November 4-8, 1997.}
Transmission and processing of (quantum) information is a physical process
in spacetime.  Therefore a formulation of 
such basic notions in quantum information theory
as the notions of composite systems, entangled states and the 
channel should include 
the spacetime variables.

In this paper entangled states  
in space and time are considered. A modification of
Bell`s equation which includes the spacetime variables is suggested.
A general relation between quantum theory
and classical theory of stochastic processes
is proposed which expresses the condition of local realism
in the form of a {\it noncommutative spectral theorem}.
Applications of this relation to the security 
of quantum key distribution
in quantum cryptography are considered.

Entangled states, i.e. the states of two particles
with the wave function which is not a product of the wave functions
of single particles,  have been studied in many 
theoretical and experimental
works starting from works of Einstein, Podolsky and Rosen, Bohm and Bell,
see e.g.~\cite{AfrSel}.

Bell's theorem~\cite{Bel} states that there are quantum
spin correlation functions that can not be represented as classical
correlation functions of separated  random variables. It has been
interpreted as incompatibility of the requirement of locality with
the statistical predictions of quantum mechanics~\cite{Bel}. For a
recent discussion of Bell's theorem see, for example
~\cite{AfrSel} - ~\cite{Vol1} and references therein.
 It is now widely accepted, as a result of
Bell's theorem and related experiments, that 
"Einstein`s local realism" must
be rejected.

Let us note however that, evidently,  the very 
formulation of the problem of locality in
quantum mechanics is based on ascribing a special role to the
position in ordinary three-dimensional space. It is rather
surprising therefore that the space dependence of the wave
function is neglected in   discussions of the problem of locality
in relation to Bell's inequalities. Actually it is the space part
of the wave function which is relevant to the consideration of the
problem of locality.

We know that
the wave function of particle includes not only the spin
part but also the part depending on spacetime variables. 
Recently it was  pointed 
out \cite{Vol1} that in fact the spacetime part of the wave 
function was neglected in the
proof of Bell's theorem. However just the spacetime part is 
crucial for considerations of property of locality of 
quantum system. Actually the spacetime part leads to an extra 
factor in quantum correlations and as a result the ordinary 
proof of Bell's theorem fails in this
case. We present a modification of Bell`s equation which
includes space and time variables.

We present a criterion of locality (or nonlocality) of quantum theory
in a realist model of hidden variables. We argue that predictions
of quantum mechanics can be consistent with   Bell's inequalities
for some Gaussian wave functions and hence Einstein's local realism is
restored in this case. 
Moreover we show that due to the expansion of the wave packet
the locality criterion 
is always satisfied for nonrelativistic
particles if regions of detectors are far enough
from each other. This result has applications to the security
of certain quantum cryptographic protocols.

We will consider  an important connection between quantum mechanics
and theory of classical stochastic processes. Consider 
for example an equation
$$
\cos(\alpha - \beta) = E\xi_{\alpha}\eta_{\beta}
$$
where $\xi_{\alpha}$ and $\eta_{\beta}$ are two random
processes \cite{Hid} and $E$ is the expectation. Bell`s theorem states
that there exists no solution of the equation for bounded
stochastic processes 
such that $|\xi_{\alpha}|\leq 1$,~
$|\eta_{\beta}|\leq 1$. 

The function $\cos(\alpha - \beta)$ 
describes the quantum mechanical correlation of
spins of two entangled particles. 
It was shown in \cite{Vol1} that if one takes into account 
the space part of the
wave function then  the quantum correlation 
in the simplest case will take
the form $g \cos (\alpha - \beta)$ instead of just  $\cos
(\alpha - \beta)$ where the parameter $g$ describes the
location of the system in space and time. In this case 
one  gets a modified equation
$$
g\cos(\alpha - \beta)= E\xi_{\alpha}\eta_{\beta}
$$
One can prove (see below) that if $g$ is small enough 
then there exists a solution of the modified equation. 

It is important  to study also a more general question:
which class of functions $f(s,t)$ admits a representation
of the form
$$
f(s,t)=Ex_sy_t
$$
where $x_s$ and $y_t$ are  bounded stochastic processes
and also analogous question for the functions of several variables
$f(t_1,...,t_n).$

Such considerations could provide a {\it noncommutative}
generalization of von Neumann`s spectral theorem.

Bell's theorem constitutes an important part in quantum
cryptography~\cite{Eke}.
It is now generally accepted that techniques of quantum
cryptography can allow secure communications between distant
parties  ~\cite{Wie} - ~\cite{VV2}. The promise of secure
cryptographic quantum key distribution schemes is based on the use
of  quantum entanglement in the spin space and  on quantum
no-cloning theorem. An important contribution of quantum
cryptography is a mechanism for detecting eavesdropping.

However in certain current
quantum cryptography protocols the space part of the wave function
is neglected. But just the space part of the wave function
describes the behaviour of particles in ordinary real
three-dimensional
 space. As a result such schemes can be secure against
eavesdropping attacks in the abstract spin space but could  be
insecure in the real three-dimensional space.
We will discuss how one can
try to improve the security of quantum 
cryptography schemes in space  by
using  a special preparation of 
the space part of the wave function,
see \cite{Vol2}.

Spacetime description is important for quantum computation
~\cite{Llo}. Some problems of quantum teleportation in space
have been discussed in ~\cite{FO}.
 
\section{Bell's Theorem}

\subsection{Bell`s Theorem and Stochastic Processes}

In the  presentation of Bell's theorem we will 
follow ~\cite{Vol1} where one
can find also more references. Bell's
theorem reads:
\begin{equation}
\cos(\alpha - \beta)\neq E\xi_{\alpha}\eta_{\beta}
\label{eq:cos}
\end{equation}
where $\xi_{\alpha}$ and $\eta_{\beta}$ are two random
processes such that $|\xi_{\alpha}|\leq 1$,~
$|\eta_{\beta}|\leq 1$ and $E$ is the expectation.
In more details:

{\bf Theorem 1.} There exists no probability space 
$(\Lambda, {\cal F}, d\rho (\lambda))$ and a pair of stochastic processes
$\xi_{\alpha}=\xi_{\alpha}(\lambda), ~\eta_{\beta}=\eta_{\beta}(\lambda),
~0\leq \alpha,\beta \leq 2\pi$ which obey 
$|\xi_{\alpha}(\lambda)|\leq 1$,~
$|\eta_{\beta}(\lambda)|\leq 1$ such that the following equation is valid
\begin{equation}
\cos(\alpha - \beta) = E\xi_{\alpha}\eta_{\beta}
\label{eq:cosin}
\end{equation}
for all $\alpha$ and $\beta$.

Here $\Lambda$ is a set, ${\cal F}$ is a sigma-algebra of subsets
and $d\rho (\lambda)$ is a probability measure, i.e. $d\rho (\lambda)
\geq 0,~\int d\rho (\lambda)=1.$
The expectation is
$$
E\xi_{\alpha}\eta_{\beta}=\int_{\Lambda}\xi_{\alpha}(\lambda)
\eta_{\beta}(\lambda)d\rho (\lambda)
$$
One can write Eq.~(\ref{eq:cosin}) as an integral equation
\begin{equation}
\cos(\alpha - \beta)=\int_{\Lambda}\xi_{\alpha}(\lambda)
\eta_{\beta}(\lambda)d\rho (\lambda)
\label{eq:integ}
\end{equation}
We say that the integral equation  (\ref{eq:integ}) has no solutions 
$(\Lambda, {\cal F}, d\rho (\lambda), 
\xi_{\alpha}, \eta_{\beta})$ with the bound
$|\xi_{\alpha}|\leq 1$,~
$|\eta_{\beta}|\leq 1.$

We will prove the theorem below. Let us
discuss now the physical interpretation of this result.

Consider a pair of spin one-half particles 
formed in the singlet spin state
and moving freely towards two detectors.  If one neglects
the space part of the wave function  then one has 
the Hilbert space $\mathbb{C}^2\otimes \mathbb{C}^2$ 
and  the quantum mechanical
correlation of two spins in the singlet state $\psi_{spin}\in
\mathbb{C}^2\otimes \mathbb{C}^2$ is
\begin{equation}
 D_{spin}(a,b)=\left<\psi_{spin}|\sigma\cdot a \otimes\sigma\cdot
b|\psi_{spin}\right>=-a\cdot b
\label{eq:eqn1}
\end{equation}
Here $a=(a_1,a_2,a_3)$ and $b=(b_1,b_2,b_3)$ 
are two unit vectors in three-dimensional space $\mathbb{R}^3$,
$\sigma=(\sigma_1,\sigma_2,\sigma_3)$ are the Pauli matrices,
$$
\sigma_1=\left(
    \begin{array}{cc}0 & 1\\ 1 & 0
    \end{array}
\right),~~
\sigma_2=\left(
    \begin{array}{cc}0 & -i\\ i & 0
    \end{array}
\right),~~
\sigma_3=\left(
    \begin{array}{cc}1 & 0\\ 0 & -1
    \end{array}
\right),
$$
 $$
 \sigma\cdot a =\sum_{i=1}^{3}\sigma_i a_i
 $$
 and
$$
\psi_{spin}=\frac{1}{\sqrt 2}
\left(\left(
    \begin{array}{c}0\\1
    \end{array}
    \right)
\otimes \left(
    \begin{array}{c}1\\
    0\end{array}
    \right)
-\left(
    \begin{array}{c}1\\
    0\end{array}
    \right)
\otimes
\left(
    \begin{array}{c}0\\
    1\end{array}
    \right)
\right)
$$
If the vectors $a$ and $b$ belong to the same plane
then one can write $-a\cdot b=\cos (\alpha - \beta)$
and hence Bell's theorem states that the function $ D_{spin}(a,b)$
Eq.~(\ref{eq:eqn1}) can not be represented in the form
\begin{equation}
\label{eq:eqn2} P(a,b)=\int \xi (a,\lambda) \eta (b,\lambda)
d\rho(\lambda)
\end{equation}
i.e.
\begin{equation}
\label{eq:Ab}
D_{spin}(a,b)\neq P(a,b)
\end{equation}
Here $ \xi (a,\lambda)$ and $  \eta(b,\lambda)$ are random  fields on the
sphere, $|\xi (a,\lambda)|\leq 1$,~  $ | \eta (b,\lambda)|\leq 1$ and
$d\rho(\lambda)$ is a positive probability measure,  $ \int
d\rho(\lambda)=1$. The parameters $\lambda$ are interpreted as hidden
variables in a realist theory. It is clear that Eq.~(\ref{eq:Ab}) can be
reduced to Eq.~(\ref{eq:cos}).

\subsection{CHSH Inequality}

To prove Theorem 1 we will use the following

{\bf Theorem 2.} Let $f_1,~f_2,~g_1$ and $g_2$ be random variables
(i.e. measured functions) on the probability space $(\Lambda, {\cal F}, d\rho (\lambda))$
such that 
$$
|f_i(\lambda)g_j(\lambda)|\leq 1,~~i,j=1,2.
$$
Denote
$$
P_{ij}=Ef_ig_j,~~i,j=1,2.
$$
Then
$$
|P_{11}-P_{12}|+|P_{21}+P_{22}|\leq 2.
$$
{\bf Proof of Theorem 2.} One has
$$
P_{11}-P_{12}=Ef_1g_1-Ef_1g_2=E(f_1g_1(1\pm f_2g_2))-
E(f_1g_2(1\pm f_2g_1))
$$
Hence
$$
|P_{11}-P_{12}|\leq E(1\pm f_2g_2)+
E(1\pm f_2g_1)=2\pm (P_{22}+P_{21})
$$
Now let us note that if $x$ and $y$ are two real numbers then
$$
|x|\leq 2\pm y~~\to~~|x|+|y|\leq 2.
$$
Therefore  taking $x=P_{11}-P_{12}$ and $y=P_{22}+P_{21}$
one gets the bound
$$
|P_{11}-P_{12}|+|P_{21}+P_{22}|\leq 2.
$$
The theorem is proved. 

The last inequality is called
the Clauser-Horn-Shimony-Holt (CHSH) inequality.
By using notations of Eq.~(\ref{eq:eqn2}) one has
\begin{equation}
\label{eq:eqn3}
 |P(a,b)-P(a,b')|+|P(a',b)+P(a',b')|\leq 2
\end{equation}
for any four unit vectors $a,b,a',b'$.

{\bf Proof of Theorem 1.} Let us denote
$$
f_i(\lambda)=\xi_{\alpha_i}(\lambda),~~
g_j(\lambda)=\eta_{\beta_j}(\lambda),~~i,j=1,2
$$
for some $\alpha_i,\beta_j.$ If one would have
$$
\cos (\alpha_i - \beta_j)=Ef_ig_j
$$
then due to Theorem 2 one should have
$$
|\cos (\alpha_1 - \beta_1)-\cos (\alpha_1 - \beta_2)|
+|\cos (\alpha_2 - \beta_1)+\cos (\alpha_2 - \beta_2)|\leq 2.
$$
However  for $\alpha_1=\pi/2,~\alpha_2=0,~\beta_1=\pi/4,
\beta_2=-\pi/4$ we obtain
$$
|\cos (\alpha_1 - \beta_1)-\cos (\alpha_1 - \beta_2)|
+|\cos (\alpha_2 - \beta_1)+\cos (\alpha_2 - \beta_2)|=2\sqrt 2
$$
which is greater than 2.
This contradiction proves Theorem 1.

It will be shown below that if one takes 
into account the space part of the
wave function then  the quantum correlation 
in the simplest case will take
the form $g \cos (\alpha - \beta)$ instead of just  $\cos
(\alpha - \beta)$ where the parameter $g$ describes the
location of the system in space and time. In this case one can get a
representation
\begin{equation}
g\cos(\alpha - \beta)= E\xi_{\alpha}\eta_{\beta}
\label{eq:gcos}
\end{equation}
if $g$ is small enough. The factor $g$ gives a contribution to
visibility or efficiency of detectors that are used in the phenomenological
description of detectors.

\section {Localized Detectors}

\subsection{Modified Bell`s equation}

In the previous section the space part of the wave function of the
particles was neglected. However exactly the space part is relevant to the
discussion of locality. The Hilbert space assigned to
one particle with spin 1/2 is  $\mathbb{C}^2\otimes L^2(\mathbb{R}^3)$
and the Hilbert space of two particles is
$\mathbb{C}^2\otimes L^2(\mathbb{R}^3)\otimes 
\mathbb{C}^2\otimes L^2(\mathbb{R}^3).$
The complete wave function is $\psi
=(\psi_{\alpha\beta}({\bf r}_1,{\bf r}_2,t))$ where $\alpha$ and
$\beta $ are spinor indices, $t$ is time  and ${\bf r}_1$ and ${\bf r}_2$
are vectors in three-dimensional space.

We suppose that there are two  detectors ("Alice" and "Bob")
which are located in space $\mathbb{R}^3$ within the
two localized regions ${\cal O}_A$ and ${\cal O}_B$ respectively, well
separated from one another. One could apply here an approach
of local algebras \cite{Bog}.

Quantum correlation describing the measurements of spins by Alice and Bob
at their localized  detectors is

\begin{equation}
\label{eq:eqn6}
 G(a,{\cal O}_A,b,{\cal O}_B)=\left<\psi| \sigma\cdot a   P_{{\cal O}_A}
 \otimes  \sigma\cdot b  P_{{\cal O}_B} |\psi\right>
\end{equation}
Here $P_{{\cal O}}$ is the projection operator onto the region ${\cal O}$.

Let us consider the case when the wave function has the form of the product
of the spin function and the space function $\psi=\psi_{spin}\phi({\bf
r}_1,{\bf r}_2)$. Then one has
\begin{equation}
\label{eq:eqn7}
 G(a,{\cal O}_A,b,{\cal O}_B)=g ({\cal O}_A,{\cal O}_B)
  D_{spin}(a,b)
\end{equation}
where the function
\begin{equation}
\label{eq:eqn8}
 g ({\cal O}_A,{\cal O}_B)=\int_{{\cal O}_A \times {\cal O}_B}|\phi({\bf
 r}_1,{\bf
 r}_2)|^2 d{\bf r}_1d{\bf r}_2
\end{equation}
describes correlation of particles in space. It is the probability to find
one particle in the region ${\cal O}_A$ and another particle in the region
${\cal O}_B$.

One has
\begin{equation}
\label{eq:eqn8g} 0\leq g ({\cal O}_A,{\cal O}_B)\leq 1
\end{equation}

{\bf Remark 1.} In relativistic quantum field theory there is no nonzero
strictly localized projection operator that annihilates the vacuum. It is
a consequence of the Reeh-Schlieder theorem.  Therefore, apparently, the
function $ g ({\cal O}_A,{\cal O}_B)$ should be always strictly smaller
than 1. 

Now one inquires whether one can write the representation
\begin{equation}
\label{eq:eqn9}
 g({\cal O}_A,{\cal O}_B)D_{spin}(a,b)=\int \xi (a,{\cal O}_A,\lambda)
 \eta (b,{\cal O}_B,\lambda) d\rho(\lambda)
\end{equation}

Note that if we are interested in the conditional probability of
finding the projection of spin along  vector $a$ for the particle
1  in the region ${\cal O}_A$ and the projection of spin along the
vector $b$ for the particle 2 in the region ${\cal O}_B$   then we
have to divide both sides of Eq.~(\ref{eq:eqn9}) by $g({\cal
O}_A,{\cal O}_B)$.

Instead of Eq~(\ref{eq:cosin}) in Theorem 1 now we have the modified
equation
\begin{equation}
g\cos(\alpha - \beta) = E\xi_{\alpha}\eta_{\beta}
\label{eq:cosinus}
\end{equation}
The factor $g$ is important. In particular one can write the following
representation \cite{VV} for $0\leq g\leq 1/2$:
\begin{equation}
\label{eq:gek}
g\cos(\alpha-\beta)=
\int_0^{2\pi}\sqrt {2g}\cos(\alpha-\lambda) \sqrt {2g}\cos(\beta-\lambda)
 \frac{d\lambda}{2\pi}
\end{equation}
Therefore if $0\leq g\leq 1/2$ then there exists a solution
of Eq.~(\ref{eq:cosinus}) where
$$
\xi_{\alpha}(\lambda)=\sqrt {2g}\cos(\alpha-\lambda),~
~\eta_{\beta}(\lambda)=\sqrt {2g}\cos(\beta-\lambda)
$$
and $|\xi_{\alpha}|\leq 1,~|\eta_{\beta}|\leq 1.$
If $g>1/\sqrt 2$ then it follows from Theorem 2 that there is no
solution to Eq.~(\ref{eq:cosinus}). We have obtained

{\bf Theorem 3.} If $g>1/\sqrt 2$ then there is no
solution 
$(\Lambda, {\cal F}, d\rho (\lambda), 
\xi_{\alpha}, \eta_{\beta})$ to Eq.~(\ref{eq:cosinus}) with the bounds
$|\xi_{\alpha}|\leq 1$,~
$|\eta_{\beta}|\leq 1.$ If $0\leq g\leq 1/2$ then there exists a solution
to Eq.~(\ref{eq:cosinus}) with the bounds
$|\xi_{\alpha}|\leq 1$,~
$|\eta_{\beta}|\leq 1.$

{\bf Remark 2.} Further results on solutions of the modified
equation have been obtained by A.K. Guschchin,
S. V. Bochkarev and D. Prokhorenko.
Local variable models for inefficient detectors are presented in
\cite{San, Lar}.

{\bf Remark 3.} A local modified equation reads
$$
|\phi (r_1,r_2,t)|^2\cos(\alpha - \beta)=E\xi (\alpha,r_1,t)
\eta (\beta,r_2,t).
$$

 \subsection {Relativistic Particles}
 
We can not immediately apply the previous considerations
to the case of relativistic particles such as photons
and the Dirac particles because in these cases
 the wave function can not be represented 
as a product of the spin part and the spacetime part.
Let us show that the wave function of photon can not
be represented in the product form. Let $A_i(k)$ be the wave
function of photon, where $i=1,2,3$ and $k\in \mathbb{R}^3$.
One has the gauge condition $k^iA_i(k)=0$ \cite{AB}. If one supposes
that the wave function has a product form $A_i(k)=\phi_i f(k)$
then from the gauge condition one gets $A_i(k)=0$.
Therefore the case of relativistic particles requires
a separate investigation. 

\subsection{Noncommutative Spectral Theory and Local Realism}

As a generalisation of the previous discussion
we would like to suggest here
a general relation between quantum theory
and  theory of classical stochastic processes~\cite{Hid}
 which expresses the condition of local realism.
Let ${\cal H}$ be a Hilbert space, $\rho $ is the density
operator, $\{A_{\alpha}\}$ is a family of self-adjoint
operators in ${\cal H}$. One says that the family
of observables $\{A_{\alpha}\}$ and the state $ \rho$
satisfy to {\it the condition of local realism} if there exists
a probability space $(\Lambda, {\cal F}, d\rho (\lambda))$
and a family of random variables $\{\xi_{\alpha}\}$ such
that  the range of $\xi_{\alpha}$ belongs to the spectrum
of $A_{\alpha}$ and for
any subset $\{A_i\}$ of mutually commutative operators
one has a representation
$$
Tr( \rho A_{i_1}...A_{i_n})=E\xi_{i_1}...\xi_{i_n}
$$  
The physical meaning of the representation is that it 
describes the quantum-classical correspondence.
If the family $\{A_{\alpha}\}$ would be a maximal commutative
family of self-adjoint operators then for pure states
the previous representation
can be reduced to the von Neumann spectral
theorem~\cite{Nai}. In our case the family $\{A_{\alpha}\}$ 
consists  from not necessary commuting operators.
Hence we will call such a representation a {\it noncommutative spectral 
representation}.  Of course one has a question for which families
of operators and states a {\it noncommutative
spectral theorem}  is valid, i.e. when we can write 
the noncommutative spectral representation. We need a noncommutative
generalization of von Neumann`s spectral theorem.

It would be helpful
to study the following problem: describe the class of
functions $f(t_1,...,t_n)$ which admits
the representation of the form
$$
f(t_1,...,t_n)=Ex_{t_1}...z_{t_n}
$$
where $x_t,...,z_t$ are random processes which obey the bounds
$|x_t|\leq 1,...,|z_t|\leq 1$. 

From the previous discussion we know that there are such 
families of operators and such states which do not 
admit the noncommutative spectral representation
and therefore they do not satisfy
the condition of local realism. Indeed let us take
the Hilbert space ${\cal H}=\mathbb{C}^2\otimes \mathbb{C}^2$
and four operators $A_1,A_2,A_3,A_4$ of the form (we denote
$A_3=B_1, A_4=B_2$)
$$
A_1=\left(
    \begin{array}{cc}\sin\alpha_1 & \cos\alpha_1
	\\ \cos\alpha_1 & -\sin\alpha_1
    \end{array}
\right)\otimes I,~~A_2=\left(
    \begin{array}{cc}\sin\alpha_2 & \cos\alpha_2
	\\ \cos\alpha_2 & -\sin\alpha_2
    \end{array}
\right)\otimes I
$$
and
$$
B_1=I\otimes\left(
    \begin{array}{cc}-\sin\beta_1 & -\cos\beta_1
	\\ -\cos\beta_1 & \sin\beta_1
    \end{array}
\right),~~B_2=I\otimes\left(
    \begin{array}{cc}-\sin\beta_2 & -\cos\beta_2
	\\ -\cos\beta_2 & \sin\beta_2
    \end{array}\right)
$$
Here operators $A_i$ correspond to operators $\sigma\cdot a$
and operators $B_i$ corresponds to operators $\sigma\cdot b$
where $a=(\cos\alpha,0,\sin\alpha),~b=(-\cos\beta,0,-\sin\beta).$
Operators $A_i$ commute with operators 
$B_j,~~[A_i,B_j]=0,~i,j=1,2$ and one has
$$
\left<\psi_{spin}|A_iB_j|\psi_{spin}\right>
=\cos(\alpha_i - \beta_j),~~i,j=1,2
$$
We know from Theorem 2 that this function can not be represented
as the expected value $E\xi_i\eta_j$  of random variables
with the bounds $|\xi_{i}|\leq 1$,~
$|\eta_{j}|\leq 1.$
 
However, as it was discussed above, the space part of the wave
function
was neglected in the previous consideration.
We suggest that in physics one could prepare
only such states and observables which satisfy the condition
of local realism. Perhaps we should restrict ourself
in this proposal to the consideration of  
only such families of observables
which satisfy the condition of relativistic local causality.
If there are physical phenomena which do not satisfy this proposal
then it would be important to {\it describe quantum processes
which satisfy the above formulated condition of local realism 
and also processes
which do not satisfy this condition}.

Let us now apply these considerations to quantum cryptography.

\section {The  Quantum Key Distribution}

\subsection{Protocol}

There are quantum cryptographic protocols with one and with
two particles, for a review see for example \cite{VV2}.
Here we shall consider the quantum key distribution
with two particles.
Ekert \cite{Eke} showed that one can use the Einstein-Podolsky-Rosen 
correlations to establish
a secret random key between two parties ("Alice" and "Bob"). Bell's
inequalities are used to check the presence of an intermediate eavesdropper
("Eve"). 
 There are two stages to the quantum cryptographic protocol, the first stage over a
quantum channel, the second over a public channel.

The quantum channel consists of a source that emits pairs of spin one-half
particles, in a singlet state. The particles fly apart towards Alice and
Bob, who, after the particles have separated, perform measurements on spin
components along one of three directions, given by unit vectors $a$ and
$b$. In the second stage Alice and Bob communicate over a public
channel.They announce in public the orientation of the detectors they have
chosen for particular measurements. Then they divide the measurement
results into two separate groups: a first group for which they used
different orientation of  the detectors, and a second group for which they
used the same orientation of the detectors. Now Alice and Bob can reveal
publicly the results they obtained but within the first group of
measurements only. This allows them, by using Bell's inequality, to
establish the presence of an eavesdropper (Eve). The results of the second
group of measurements can be converted into a secret key. One supposes that
Eve  has a detector which is located within the region ${\cal O}_E$ and she
is described by hidden variables $\lambda$.

We will interpret Eve as  a  hidden variable in a realist theory and  will
study whether the quantum correlation  can be
represented in the form Eq.~(\ref{eq:eqn9}).  From  Theorem 3
 one can see that if the following
inequality
\begin{equation}
\label{eq:eqn10}
 g ({\cal O}_A,{\cal O}_B)\leq 1/ 2
\end{equation}
is valid for  regions  ${\cal O}_A$ and ${\cal O}_B$ which are well
separated from one another then there is no violation of the CHSH
inequalities (\ref{eq:eqn3}) and therefore Alice and Bob can not detect the
presence of an eavesdropper. On the other side, if for a pair of well
separated regions ${\cal O}_A$ and ${\cal O}_B$ one has
\begin{equation}
\label{eq:eqn11}
 g ({\cal O}_A,{\cal O}_B) > 1/\sqrt 2
\end{equation}
then it could be a  violation of the realist locality in these regions for
a given state. Then, in principle, one can hope to detect an eavesdropper
in these circumstances.

Note that if we set $ g({\cal O}_A,{\cal O}_B)=1$ in (\ref{eq:eqn9}) as it
was done in the original proof of Bell's theorem, then it means we did a
special preparation of the states of particles to be completely localized
inside of detectors. There exist such well localized states (see however
the previous Remark) but there exist also another states, with the wave
functions which are not very well localized inside the detectors, and still
particles in such states are also observed in detectors. The fact that a
particle is observed inside the detector does not mean, of course, that its
wave function is strictly localized inside the detector before the
measurement.  Actually one has  to perform a thorough investigation of the
preparation and the  evolution of our entangled states in space and time if
one needs to estimate the function $ g({\cal O}_A,{\cal O}_B)$.

\subsection {Gaussian Wave Functions}

Now let us consider the criterion of locality for Gaussian wave
functions. We will show that with a reasonable accuracy there is
no violation of locality in this case. Let us take the wave
function $\phi$ of the form $\phi=\psi_{1}({\bf r}_1)\psi_{2}({\bf
r}_2)$  where the individual wave functions have the moduli
\begin{equation}
\label{eq:eqn12}
|\psi_{1}({\bf
 r})|^2 =(\frac{m^2}{2\pi})^{3/2}e^{-m^2{\bf r}^2/2},~~
|\psi_{2}({\bf
 r})|^2 =(\frac{m^2}{2\pi})^{3/2}e^{-m^2 ({\bf r}- {\bf l})^2/2}
\end{equation}
We suppose that  the length of the vector ${\bf l}$ is much larger
than $1/m$. We can make measurements of  $P_{{\cal O}_A}$ and
$P_{{\cal O}_B}$ for any  well separated regions  ${\cal O}_A$ and
${\cal O}_B$. Let us suppose a rather nonfavorite case for the
criterion of locality when the wave  functions of the particles
are almost localized inside the regions ${\cal O}_A$ and ${\cal
O}_B$ respectively. In such  a case the function $g ({\cal
O}_A,{\cal O}_B)$ can take values near its maximum. We suppose
that the region ${\cal O}_A$ is given by $|r_i|<1/m, {\bf
r}=(r_1,r_2,r_3) $ and the region ${\cal O}_B$ is obtained from
${\cal O}_A$ by translation on ${\bf l}$. Hence $\psi_{1}({\bf
r}_1)$ is a Gaussian function with modules appreciably different
from zero only in ${\cal O}_A$ and similarly $\psi_{2}({\bf r}_2)$
is localized in the region ${\cal O}_B$. Then we have
\begin{equation}
\label{eq:eqn13}
 g ({\cal O}_A,{\cal O}_B)=\left(\frac{1}{\sqrt {2\pi}}\int_{-1}^1
e^{-x^2/2}dx\right)^6
\end{equation}
One can estimate (\ref{eq:eqn13}) as
\begin{equation}
\label{eq:eqn14}
 g ({\cal O}_A,{\cal O}_B)< \left(\frac{2}{\pi}\right)^3
\end{equation}
which is smaller than $1/2$. Therefore the locality criterion
(\ref{eq:eqn10}) is satisfied in this case.

\subsection{Expansion of Wave Packet}

Let us remind that there is a well known effect of expansion of
wave packets due to the free time evolution. If $\epsilon$ is the
characteristic length of the Gaussian wave packet describing a
particle of mass $M$ at time $t=0$  then at time $t$ the
characteristic length  $\epsilon_t$ will be
\begin{equation}
\label{eq:epsil}
\epsilon_t=\epsilon\sqrt{1+\frac{\hbar^2t^2}{M^2\epsilon^4}}.
\end{equation}

It tends to $(\hbar/M\epsilon)t$ as $t\to\infty$. Therefore the
locality criterion is always satisfied for nonrelativistic
particles if regions ${\cal O}_A$ and ${\cal O}_B$ are far enough
from each other. 

\section {Conclusions}

We have studied some problems in quantum information theory
which requires the inclusion of spacetime variables.
In particular entangled states  
in space and time were considered. A modification of
Bell`s equation which includes the spacetime variables is suggested and 
investigated.
A general relation between quantum theory
and  theory of classical stochastic processes was proposed
 which expresses the condition of local realism
 in the form of a noncommutative spectral theorem.
  Applications of this relation to the security 
of quantum key distribution
in quantum cryptography were considered.

There are many interesting open problems in the approach
to quantum information in space and time discussed in this paper.
Some of them related with the noncommutative spectral theory and
theory of classical stochastic processes have been discussed above.

In quantum cryptography there are important open problems
which require  further investigations.
In quantum cryptographic protocols with two entangled photons
 to detect the eavesdropper's presence
by using Bell's inequality we have to estimate the function  $ g({\cal
O}_A,{\cal O}_B)$. To
increase the detectability of the eavesdropper
one has to do a thorough investigation of the
process of preparation  of the entangled state and then its  evolution in
space and time towards Alice and Bob.
One has to develop a proof of the security of such a protocol.

In the previous section Eve was interpreted as an abstract hidden variable.
However one can assume that more information about Eve is available. In
particular one can assume that she is located somewhere in space in a
region ${\cal O}_E.$ It seems one has to study a generalization of the
function  $ g({\cal O}_A,{\cal O}_B)$, which depends not only on the Alice
and Bob locations ${\cal O}_A$ and $ {\cal O}_B$ but also depends on the
Eve location $ {\cal O}_E$, and try to find a strategy which leads to an
optimal value of this function.

\section*{Acknowledgments}

I am grateful to  T. Hida and  K. Saito for
the invitation to the conference
on Quantum Information
at Meijo University, Japan and to  M. Rasetti
for the invitation to the 
ISI Workshop  on Quantum Computer Theory
 held at the ISI Foundation in Torino, Italy. 
I would like to thank A. Baranov, S. Bochkarev, A. Chebotarev,
W. De Baere, Yu. Drozzinov, L. Fedichkin, C. Fuchs, A. Guschchin,
A. Holevo, 
A. Khrennikov,  J.-A. Larsson,
S. Lloyd, Hoi-Kwong Lo, D. Mayers, D. Mermin, M. Ohya,  
K. Parthasarathy, D. Prokhorenko,  
 R. Roschchin and  L. Vaidman
for helpful discussions and remarks.
This work was supported in part by RFFI 99-0100866
 and by INTAS 99-00545 grants.

\end{document}